\documentclass[twocolumn,secnumarabic,amssymb, showpacs, nobibnotes, aps, prd]{revtex4-1}

\usepackage{graphicx}
\usepackage{subfigure}
\usepackage{ae} 
\usepackage{float}  
\usepackage{amsmath}
\usepackage{color}

\begin{document}

\title{Resonant phonon-magnon interactions in free-standing metal-ferromagnet multilayer structures}

\author{Urban Vernik $^{1}$}
\email[]{urban.vernik@polytechnique.edu}

\author{Alexey M.~ Lomonosov$^{2}$}

\author{Vladimir S.~Vlasov$^{1}$}
\author{Leonid N.~Kotov$^{1}$} 

\author{Dmitry A.~ Kuzmin$^{1}$} 
\author{Igor V.~Bychkov$^{1}$} 
\author{Paolo Vavassori$^{3,4}$}
\author{Vasily V.~ Temnov$^{1}$}
\email[]{vasily.temnov@cnrs.fr}

\affiliation{$^1$LSI, Ecole Polytechnique, CEA/DRF/IRAMIS, CNRS, Institut Polytechnique de Paris, F-91128, Palaiseau, France}
\affiliation{$^2$University of Applied Sciences, 77723 Gengenbach, Germany}
\affiliation{$^3$CIC nanoGUNE, E-20018 Donostia-San Sebastian, Spain} 
\affiliation{$^4$IKERBASQUE, Basque Foundation for Science, E-48013 Bilbao , Spain}


\date{\today}

\begin{abstract}
 We analyze resonant magneto-elastic interactions between standing perpendicular spin wave modes (exchange magnons) and longitudinal acoustic phonon modes in free-standing hybrid metal-ferromagnet bilayer and trilayer structures. Whereas the ferromagnetic layer acts as a magnetic cavity, all metal layers control the frequencies and eigenmodes of acoustic vibrations. The here proposed design allows for achieving and tuning the spectral and spatial modes overlap between phonons and magnons that results in their strong resonant interaction. Realistic simulations for gold-nickel multilayers show that sweeping the external magnetic field should allow for observing resonantly enhanced interactions between individual magnon and phonon modes in a broad range of frequencies spanning from tens of GHz up to several hundreds of GHz, which can be finely tuned through the multilayer design. Our results would enable the systematic study and the deep understanding of resonantly enhanced magneto-elastic coupling between individual phonon and magnon modes up to frequencies of great contemporary fundamental and applied interest.   
\end{abstract}

\pacs{Valid PACS appear here}


\maketitle

\section{Introduction}

Over the last decade studies of ultrafast magneto-elastic interactions in ferromagnetic thin films have attracted significant attention motivated by possibility of  control of the magnetization dynamics at the nano-scale \cite{Scherbakov2010,kim2012ultrafast,Thevenard2010,janusonis2016ultrafast,chang2017parametric,Vlasov_2022}, towards magnetization switching \cite{thevenard2013irreversible,VlasovPRB2020}. Being intrinsically weak, magneto-elastic interactions can be enhanced by quasi-monochromatic elastic excitations tuned in resonance with acoustic vibrations \cite{janusonis2016ultrafast,chang2017parametric,berk2019strongly,Vlasov_2013,Vlasov_2014}. Magneto-elastic interactions can excite not only the spatially homogeneous mode of ferromagnetic resonance (FMR) precession, but also spin waves (magnons). In case of ferromagnetic thin films of nanometer thickness perpendicular (exchange) spin wave resonances across the film \cite{van2002all} can be excited by ultrashort picosecond pulses of longitudinal acoustic phonons propagating through the film \cite{bombeck2012excitation,BesseJMMM}. In this case, the resonant enhancement of phonon-magnon interactions requires creating a magneto-acoustic cavity across the film, namely a cavity enabling the frequency/phase matching between the magnetic and acoustic modes. Such a cavity would enable the systematic study and the deep understanding of magneto-elastic interactions between individual phonon and magnon modes. A way to create a magneto-acoustic cavity is to use freestanding ferromagnetic thin films \cite{kim2017magnetization}. However, the difference between dispersion relations of phonons and magnons, in experimentally accessible nm-thick films, does not allow achieving the frequency/phase matching between magnetic and acoustic modes, apart from the trivial case of the magnetic uniform FMR precession mode. For example, a recently reported experiment in a freestanding YIG/GGG/YIG trilayer demonstrated fingerprints of the magneto-elastic coupling between two FMR-resonances in spatially separated YIG thin films, mediated by long-lived standing phonon modes \cite{an2020coherent,an2022bright}.    

In this article we propose a pathway to design magneto-acoustic cavities using free-standing metal-ferromagnet multilayers that can sustain exchange magnons and acoustic modes, and can be adjusted to achieve the frequency/phase matching between the modes to ensure their strong coupling. The proposed design allows the independent tuning of the acoustic and magnetic cavities thereby enabling the observation of magneto-elastic interactions with higher-order (exchange) magnon modes with nanometer wavelengths and high frequencies exceeding 100~GHz. Indeed, in such structures the magnon spectrum is determined just by the thickness of the ferromagnetic layer while the spectrum of acoustic vibrations is determined by the total thickness of the multilayer. Thus, an independent control and frequency/phase matching between individual phonon and magnon modes becomes possible. The present manuscript presents a detailed theoretical study, notably from the acoustic perspective, of phonon-magnon interactions in free-standing bi- and trilayers targeting the fundamental understanding of magneto-elastics with high-frequency exchange magnons  towards their experimental observation in ultrafast magneto-optical pump-probe experiments. The dominant role of elastic boundary conditions is inspected for both types of structures and advantages of trilayer structures are discussed within the framework of modes symmetry.     

The paper is organized as follows. Section II discusses basics of exchange magnons, generation of ultrashort laser-induced acoustic transients, their decomposition in acoustic eigenmodes and equations between  magneto-elastic interactions induced by  in free-standing metal/ferromagntet multilayers. Section III discusses the details of acoustic mode decomposition in bilayers and symmetric trilayers and Section IV describes the numerical procedure to decompose an arbitrary initial strain distribution in (non-orthogonal) acoustic strain eigenmodes. Section V presents the numerical results comparing resonant magneto-elastic interactions in Ni/Au bilayers and Au/Ni/Au trilayers and their interpretation. The discussion (Section VI) discusses the obtained results within the context of recent experimental works.

\section{Physical model}
\begin{figure}
	\footnotesize{} \centering
 \includegraphics[width=0.8\columnwidth]{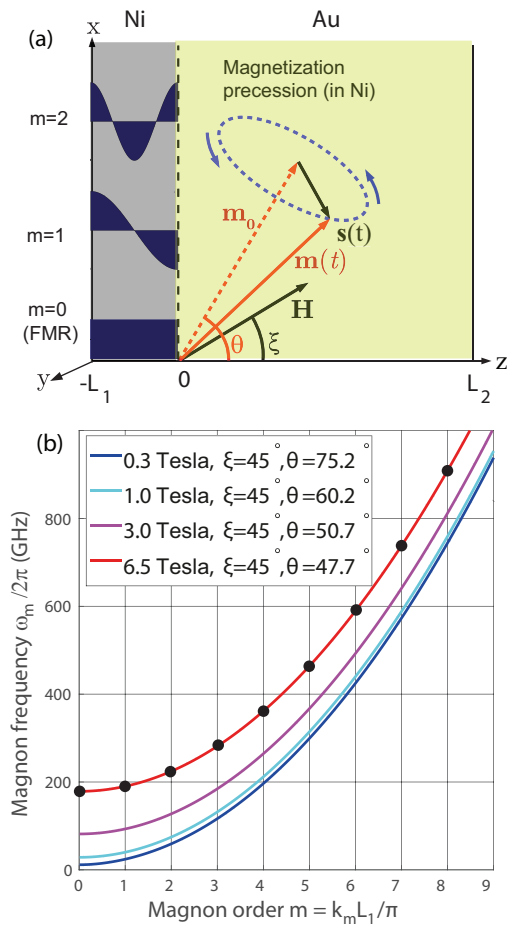}
 \caption{\label{fig1} (a) In a free-standing Ni/Au bilayer the magnetic excitations are assumed to be localized in Nickel. Standing modes of exchange magnons (with magnon index $m=0,1,2$ etc.) precess around the equilibrium magnetization direction ${\bf m_0}$ under the presence of a tilted external magnetic field ${\bf H}$. (b) Magnon frequencies $\omega_m$ depend on the magnon order $m=k_mL_1/\pi$ and the amplitude of an external magnetic field (see text for details); $L_1=30~$nm.}
\end{figure}

An example of a simple and experimentally feasible model system for studying resonant phonon-magnon interactions is a free-standing metal-ferromagnet (gold-nickel) bilayer 
Assuming that spin transport across metal-ferromagnetic interface can be neglected in this study, we are going to discuss their magnetic and acoustic properties separately, see Fig.~1 for standing spin-wave modes localized in Nickel and Fig.~2 for acoustic eigenmodes in the entire bilayer and their excitation.

In this article we focus on the analysis of the acoustically driven precessional motion of magnetisation vector of constant, time-independent unit length 
$|{\bf m}(t)|=1$ around the stationary magnetization direction $\bf{m_0}$, see Fig.~1(a).
The latter is parallel to the effective magnetic field $\mathbf{H}_{eff}={\bf H}+{\bf H_d}$, where ${\bf H}=(H\sin\xi,0,H\cos\xi)$ is the external magnetic field and ${\bf H_d}=(0,0,-M_0\cos\theta)$ is the demagnetizing field for a thin ferromagnetic film characterized by the saturation magnetization $M_0$. It is instructive to note that in this non-collinear geometry the magnetization angle $\theta$, which minimizes the magneto-static free energy density, consisting of a thin-film shape anisotropy and the Zeeman term, leads to the following relation 
\begin{equation}
\sin(\theta-\xi)=\frac{M_0}{2H}\sin 2\theta\,,
\label{fmr_equation for the angles}
\end{equation}
which is solved numerically to obtain $\theta=\theta(H,\xi)$.

The dispersion relation for spin waves has been obtained by the linearization of the Landau-Lifshitz-Gilbert equations with an additional exchange term in the vicinity of the equilibrium magnetization direction $\theta(\xi)$ \cite{BesseJMMM}. In a thin film of thickness $L_1$ the general form of the {\it spatially inhomogeneous} magnetization precession ${\bf m}(z,t)={\bf m_0} +{\bf s}(z,t)$ can be represented as a superposition in magnetic eigenmodes, i.e. the perpendicular standing spin wave resonances
\begin{equation} 
\label{mag_mode_decomposition}
    {\bf s}(z,t)=\sum_{m=0}^\infty {\bf s}^{(m)}(t)\cos(k_mz) \,,
\end{equation}
for $m=0,1,2,...$ across the film with quantized wave vectors $k_m=\pi m/L_1$ and frequencies 
\begin{widetext}
\begin{equation}
\label{fmr_magnon_tilted}
\omega_m=\gamma\mu_{0}\sqrt{(H\cos\xi-
(M_0-{\tilde D}k^2_m)\cos\theta)^2+(H\sin\xi+{\tilde D}k^2_m\sin\theta)
(H\sin\xi+(M_0+{\tilde D}k^2_m)\sin\theta)}\,,
\end{equation}
\end{widetext}
where ${\tilde D}=D^*/(\hbar\gamma\mu_0)$ is the exchange stiffness (the numerical value $D^*$=430~[meV\,{\AA}$^2$] for Ni is taken from Ref.~\cite{van2002all}), $\gamma$ stands for the gyromagnetic ratio, $\mu_0$ is vacuum permeability and $h$ is Planck's constant. Fig. 1(b) shows that the magnon frequencies depend on the magnon order $m$ and can be tuned on the amplitude $H$ and/or the angle $\xi$ of an external magnetic field.

Now we turn to the discussion of acoustic properties. Optical excitation of a free-standing absorbing thin film (we have chosen a 30~nm Ni/100~nm Au bilayer for illustration purposes) by an ultrashort, typically femtosecond, laser pulse results in ultrafast optical heating of nickel thin film \cite{saito2003picosecond,kim2017magnetization}. The subsequent thermoelastic expansion results in the generation of an ultrashort picosecond acoustic pulse $\varepsilon_{zz}(z,t)$ bouncing back and forth between nickel-air and gold-air interfaces, see Fig.~2(a). This propagating pulse can be represented as a superposition of acoustic eigenmodes: 
\begin{equation} \label{ac_mode_decomposition}
    \varepsilon_{zz}(z,t)=\sum_{n=1}^\infty a_n e^{i\omega_n t}\varepsilon_{zz}^{(n)}(z)\,,
\end{equation}
oscillating at frequencies $\omega_n$ and characterized by time-independent expansion coefficients $a_n$. Spatial profiles of the first five acoustic eigenmodes are presented in Fig.~2(b). Strain profiles in Fig.~2(a) have been obtained using Eq.~(\ref{ac_mode_decomposition}); the details of acoustic mode calculations will be provided in the next section.

After having provided this unified description of magnetic and acoustic eigenmodes we discuss their coupling. The long-lived acoustic transients, to be viewed either as propagating acoustic pulses or a superposition of individual acoustic modes, can excite the magnetization dynamics in a thin ferromagnetic layer via the magneto-strictive mechanism \cite{bombeck2012excitation,BesseJMMM}. 

\begin{figure}
	\footnotesize{} \centering
 \includegraphics[width=0.9\columnwidth]{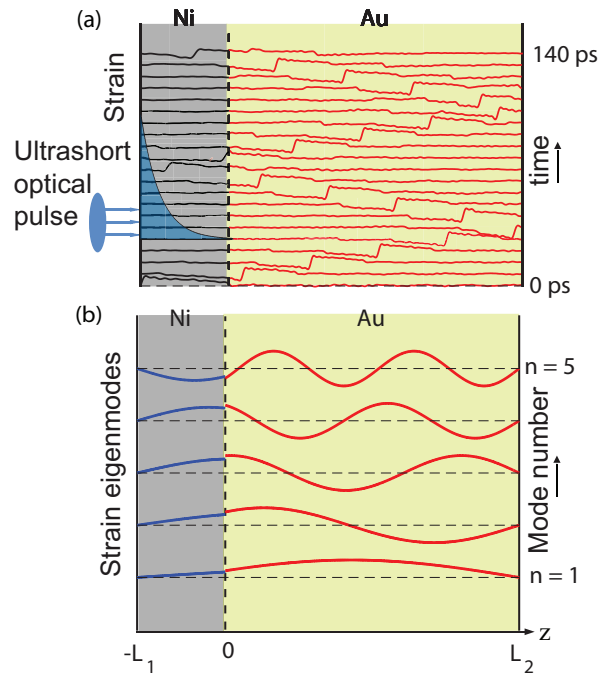}
 \caption{\label{fig2} (a) Ultrafast optical excitation of a free-standing Ni/Au bilayer generates ultrashort, picosecond strain pulses $\varepsilon_{zz}(z,t)$ bouncing back and forth in the layer. (b) The strain $\varepsilon_{zz}(z,t)$ in (a) can be represented as a superposition of acoustic eigenmodes numbered by their mode index $n=1,2,3$ etc.; $L_1=30~$nm, $L_2=70~$nm.}
\end{figure}

It has been shown in the previous work \cite{BesseJMMM} that in the linear, small-amplitude approximation, the dynamics of an individual magnon mode $s^{(m)}(t)$ can be approximated by an equation of a magneto-elastically driven harmonic oscillator:
\begin{equation}\label{magnon_oscillator}
\frac{d^2s_z^{(m)}}{dt^2}+2\alpha\omega_m\frac{ds_z^{(m)}}{dt} + \omega_m^2s^{(m)}_z = f_m(t)\,,
\end{equation}
where the damping is determined by the dimensionless Gilbert damping parameter $\alpha$ (we have assumed $\alpha=0.02$ in this work \cite{Salikhov2019}) and the parametric magneto-elastic modulation of magnon frequency \cite{chang2017parametric} is neglected. The magneto-elastic driving force is determined by a rather complex expression, a product of a prefactor and an integral term:  
\begin{equation}\label{driving}
    f_m(t)=P_m({\bf H})\int_{-L_{1}}^0\varepsilon_{zz}(z,t)\cos(k_mz)\mathrm{d}z\,,
\end{equation}
where the prefactor
\begin{equation}\label{PH}
    P_m({\bf H}) = \frac{\mu_0\gamma^2b_1\sin(2\theta)\left({\tilde D}k_m^2\sin\theta+H\sin\xi\right)}{M_0L_1}
\end{equation}   
depends on the nickel's exchange stiffness ${\tilde D}$ and magnetostrictive coefficient $b_1$. It turns to zero both for in-plane ($\theta=0$) or out-of-plane ($\theta=\pi/2$) magnetization directions. This restriction imposes the use of an experimental configuration with with the magnetization vector tilted with respect to the surface normal. 
Using the strain decomposition in eigenmodes (see Eq.~(\ref{ac_mode_decomposition})),  the integral term 
\begin{equation}\label{IntegralTerm}
    \int_{-L_{1}}^0\varepsilon_{zz}(z,t)\cos(k_mz)\mathrm{d}z=\sum_{n=1}^{\infty}a_nI_{mn}{\rm e}^{i\omega_nt}
\end{equation}
is determined by the acoustic spectrum $a_n$ and spatial overlap integrals $I_{mn}$ between individual magnetic acoustic modes: 
\begin{equation}\label{I_mn}
    I_{mn}=\int_{-L_{1}}^0\varepsilon_{zz}^{(n)}(z)\cos(k_mz)\mathrm{d}z\,.
\end{equation}
The driven solution of Eq.~(\ref{magnon_oscillator}) can be written in a compact analytical form
\begin{equation}\label{m_H_omega}
    s_z({\bf H},\omega) = \sum_{m=0}^\infty\sum_{n = 1}^{\infty}\frac{ P_m({\bf H})a_nI_{nm}}{\omega_m^2-\omega_n^2+2i\alpha\omega_m\omega_n}\,,
\end{equation}
where resonant interactions between individual magnetic and acoustic modes are described by the Lorentz functions. According to Eq.~(\ref{m_H_omega}) the magnetization oscillates at discrete frequencies $\omega=\omega_n$ of the acoustic driving force. In order to take into account for finite lifetimes of acoustic modes it is convenient to assume that acoustic lifetimes $\tau_n^{(ac)}=1/(\alpha_{ac}\omega_n)$ for individual modes are characterized by a single phenomenological damping parameter $\alpha_{ac}$, ($\alpha_{ac}=0.005$ throughout of the manuscript, the order-of-magnitude taken from \cite{hatanaka2022chip}). 
Then Eq.~(\ref{m_H_omega}) is transforms in the final expression
\begin{equation}\label{m_H_omega_AcDamping}
    s_z({\bf H},\omega) = \sum_{m=0}^\infty\sum_{n = 1}^{\infty}\frac{ P_m({\bf H})a_nI_{nm}}{\omega_m^2-\omega_n^2+2i\alpha\omega_m\omega_n}L_n^{(ac)}(\omega,\omega_n)\,,
\end{equation}
where each acoustic mode becomes broadened by the the acoustic Lorentzian  $L_n^{(ac)}(\omega,\omega_n)\propto 1/(\omega^2-\omega_n^2+i\alpha_{ac}\omega\omega_n)$. The rest of the manuscript will be devoted to the numerical analysis of this solution as well as individual terms in this equation for model bi-layer and trilayer structures.  Equation ~(\ref{m_H_omega_AcDamping}) is implemented numerically to generate smooth two-dimensional maps as a function of the continuous frequency $\omega$ and the magnetic field strength $H$ in figures 4-6.

\section{Acoustic eigenmodes in freestanding multilayers}
To calculate longitudinal acoustic eigenmodes displayed in Fig.~2(b), we represent the elastic displacement $u_{z}$ in the $z$-direction in each layer as the sum of right and left propagating plane waves, respectively:
\begin{align}
\begin{split}\label{disp}
    u_z^{(1)} &= A{\rm e}^{i(-k^{(1)}z+\omega t)} + B{\rm e}^{ i(k^{(1)}z+\omega t)}\\
    u_z^{(2)} &= C{\rm e}^{i(-k^{(2)}z+\omega t)} + D{\rm e}^{i(k^{(2)}z+\omega t)}\,
\end{split}
\end{align}
The associated strains are given by:
\begin{align} 
\begin{split}\label{strains}
    \epsilon_{zz}^{(1)} = \frac{\partial u_z^{(1)}}{\partial z}\,,\,
    \epsilon_{zz}^{(2)} = \frac{\partial u_z^{(2)}}{\partial z}\,.
\end{split}
\end{align}
The expressions (\ref{disp}) and (\ref{strains}) satisfy the following boundary conditions:
\begin{align}
\begin{split}\label{bounary}
    \epsilon_{zz}^{(1)}(-L_1) &= 0\\
    \epsilon_{zz}^{(2)}(L_2) &= 0\\
    u_z^{(1)}(0) &= u_z^{(2)}(0)\\
    v_1Z_1\epsilon_{zz}^{(1)}(0) &= v_2Z_2\epsilon_{zz}^{(2)}(0)\,,
\end{split}
\end{align}
where the acoustic impedance $Z = v\rho$ is determined by the acoustic velocity $v$ and the material density $\rho$. This linear system of four equations is quanitified by a matrix
\begin{equation}\label{matrix}
    \begin{pmatrix}
    {\rm e}^{i\omega\frac{L_1}{v_1}} & -{\rm e}^{-i\omega\frac{L_1}{v_1}} & 0 & 0\\
    1 & 1 & -1 & -1\\
    -Z_1 & Z_1 & Z_2 & -Z_2\\
    0 & 0 & {\rm e}^{-i\omega\frac{L_2}{v_2}} & -{\rm e}^{i\omega\frac{L_2}{v_2})}
\end{pmatrix}
\end{equation}
leading to the following secular equation: 
\begin{equation}\label{ac_disp_rel}
    Z_1\sin\left(\omega\frac{L_1}{v_1}\right)\cos\left(\omega\frac{L_2}{v_2}\right) + Z_2\sin\left(\omega\frac{L_2}{v_2}\right)\cos\left(\omega\frac{L_1}{v_1}\right) = 0
\end{equation}
which roots $\omega_n$ are obtained numerically. The corresponding eigenvectors $\{A_n,B_n,C_n,D_n\}$ read:   
\begin{align}
\begin{split}\label{eigenvectors}
    B_n &= A_n {\rm e}^{2i\omega_n\frac{L_1}{v_1}}\\
    C_n &= \frac{Z_1\left(1 - B_n\right)+Z_2\left(1 + B_n\right)}{2Z_2}\\
    D_n &= A_n + B_n - C_n\,,
\end{split}
\end{align}
where $A_n$ is chosen to normalize to equal acoustic energies for all modes, with the elastic energy on the $n$-th acoustic mode being:
\begin{equation}\label{elasticenergy}
    \frac{1}{2}\left(\int_{-L_{1}}^0v_{1}Z_{1}(\varepsilon^{(n)}_{zz}(z))^2\mathrm{d}z + \int_{0}^{L_{2}}v_{2}Z_{2}(\varepsilon^{(n)}_{zz}(z))^2\mathrm{d}z\right)
\end{equation}
The results of numerical calculations for bilayers can be generalized to calculate acoustic eigenmodes in symmetric metal/ferromagnet/metal trilayers in a rather straight-forward manner. Using the system of coordinates in Fig.~\ref{fig1}, a symmetric trilayer can be built by reflecting the bilayer shown in Fig.~\ref{fig1} symmetrically about its left boundary plane $z=-L_1$, so that the structure now comprises two gold layers at both sides with thickness $L_2$ and the nickel layer with thickness $2L_1$ in the center. It is convenient and reader friendly to denote the half thickness of nickel by $l_1$ and the gold layer thickness as $l_2$ for trilayers. In this new notation the plane of symmetry of the trilayer lies at $z=-l_1$, gold-nickel interface lies at $z=0$, and the right-hand free boundary of gold lies at $z=l_2$. In that symmetric trilayer there exist two types of eigenmodes of elastic vibrations, namely the symmetric ones with displacements on opposites sides of the plane of symmetry $z=-l_1$ related by $u_{+}=-u_{-}$, and antisymmetric ones with $u_{+}=u_{-}$. Due to the continuity of displacements and stresses, at the symmetry plane, the displacement $u(-l_1)$ vanishes for the symmetric modes, and its spatial derivative (and hence the stress $\varepsilon_{zz}(-l_1)$ - for the antisymmetric modes. Therefore, the set of antisymmetric eigenmodes coincides with those calculated above for freestanding bilayers if replace thicknesses $l_i$ with $L_i$. Due to the symmetry of the structure we can consider only the half of it with the boundary conditions defined at the free boundary, at the interface plane, and at the plane of symmetry. The total set ob boundary conditions for symmetric modes is different and takes the following form: 
\begin{align}\label{bc_3layer}
\begin{split}
u^{(1)}_z(-l_{1})&=0\\
\varepsilon^{(2)}_{zz}(l_{2})&=0\\
u^{(1)}_z(0)&=u^{(2)}(0)\\
v_{1}Z_{1}\varepsilon^{(1)}_{zz}(0)&=v_{2}Z_{2}\varepsilon^{(2)}_{zz}(0)\,,
\end{split}
\end{align}
for the symmetric modes, and for the antisymmetric ones the first condition in \eqref{bc_3layer} should be replaced with $u^{(1)}_z(-l_{1})~=0$. Note that the antisymmetric modes in trilayer are equivalent to the bilayer modes (\eqref{bounary}), hence \eqref{matrix} remains valid for the antisymmetric modes with substitution $L_1=2l_1$.
For the symmetric modes the matrix of boundary conditions differs in only signs of two elements:
\begin{equation} \label{antis_matrix}
    \begin{pmatrix}
    {\rm e}^{-i\omega\frac{l_1}{v_1}} & {\rm e}^{i\omega\frac{l_1}{v_1}} & 0 & 0\\
    0 & 0 & {\rm e}^{i\omega\frac{l_2}{v_2}} & -{\rm e}^{-i\omega\frac{l_2}{v_2}}\\
    1 & 1 & -1 & -1\\
    -Z_1 & Z_1 & Z_2 & -Z_2
\end{pmatrix}
\end{equation}
Eigenfrequencies of the symmetric trilayer system are the solutions to the secular equation which make the determinant of the boundary condition matrix vanish. It can be written in a concise form for both types of modes:
\begin{equation}\label{DiepersionSymmetricAntysymmetric}
\tan\left( \omega\frac{l_{1}}{v_{1}}+\alpha \right) \tan\left( \omega\frac{l_{2}}{v_{2}}\right)=\frac{Z_{1}}{Z_{2}}
\end{equation}
with phase parameter $\alpha=0$ for the symmetric modes and $\alpha=\pi/2$ for the antisymmetric ones.

\section{Decomposition of an arbitrary input strain into acoustic modes}

Using an example of a bilayer here we show how to decompose any spatio-temporal strain distribution $\varepsilon_{zz}(z,t)$ in acoustic eigenmodes:
\begin{equation}
    \varepsilon_{zz}(z,t)=\sum_{n=1}^\infty a_n e^{i\omega_n t}\varepsilon_{zz}^{(n)}(z),
\end{equation}
where every $a_n$ satisfy the following algebraic matrix equation:
\begin{equation}
    \int_{-L_1}^{L_2}\varepsilon(z,t=0)\varepsilon_{zz}^{(n)}(z)\mathrm{d}z = \sum_{n'}a_{n'}\int_{-L_1}^{L_2}\varepsilon_{zz}^{(n')}(z)\varepsilon_{zz}^{(n)}\mathrm{d}z\,.
\end{equation}
Since the acoustic eigenmodes in multilayer structures above are in general not orthogonal the coefficients $a_n$s should obey the following matrix equation:
\begin{equation}
    \hat{M}\Vec{a} = \Vec{b},
\end{equation}
where
\begin{equation}
    (\hat{M})_{ij} = \int_{-L_1}^{L_2}\varepsilon_{zz}^{(i)}(z)\varepsilon_{zz}^{(j)}(z)\mathrm{d}z
\end{equation}
and
\begin{align}
    &\Vec{a} =
    \begin{pmatrix}
    a_1\\
    \vdots\\
    a_{cutoff}
    \end{pmatrix}
    &\Vec{b} = 
    \begin{pmatrix}
    \int_{-L_1}^{L_2}\varepsilon(z,t=0)\varepsilon_{zz}^{(1)}(z)\mathrm{d}z\\
    \vdots\\
    \int_{-L_1}^{L_2}\varepsilon(z,t=0)\varepsilon_{zz}^{(cutoff)}(z)\mathrm{d}z
    \end{pmatrix}\,.
\end{align}
By numerical inversion of a truncated matrix $\hat{M})$, with the cutoff-index being chosen such that the results do not depend on it,  delivers the desired expansion coefficients (or the acoustic spectrum) $a_n$.

The performance of this decomposition method is illustrated in Fig.~3, where the initial strain profile in Ni (exponential with a characteristic decay length of approximately 20~nm \cite{saito2003picosecond}) is accurately reproduced as a superposition of the first 220 acoustic eigenmodes. This initial strain distribution, i.e. exponential in Ni and zero in Au, will be used in all numerical calculations in the forthcoming sections.

\begin{figure}
	\footnotesize{} \centering
 \includegraphics[width=1.0\columnwidth]{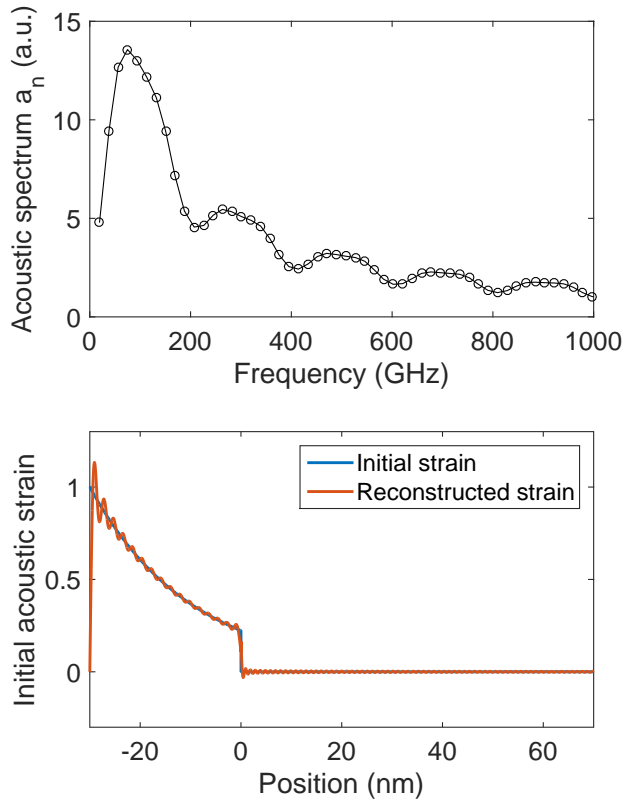}
 \caption{\label{Fig3_new} (a) The acoustic spectrum $a_n$ for a 30nm Ni/70nm Au bilayer obtained by using the matrix inversion protocol. (b) The reconstructed pulse with $n_{cutoff}$=220 allows for accurate reconstruction of the initial exponential strain profile.}
\end{figure}

\section{Resonant magneto-elastic coupling in multilayers}

\begin{figure*} 
	\footnotesize{} \centering
 \includegraphics[width=2.0\columnwidth]{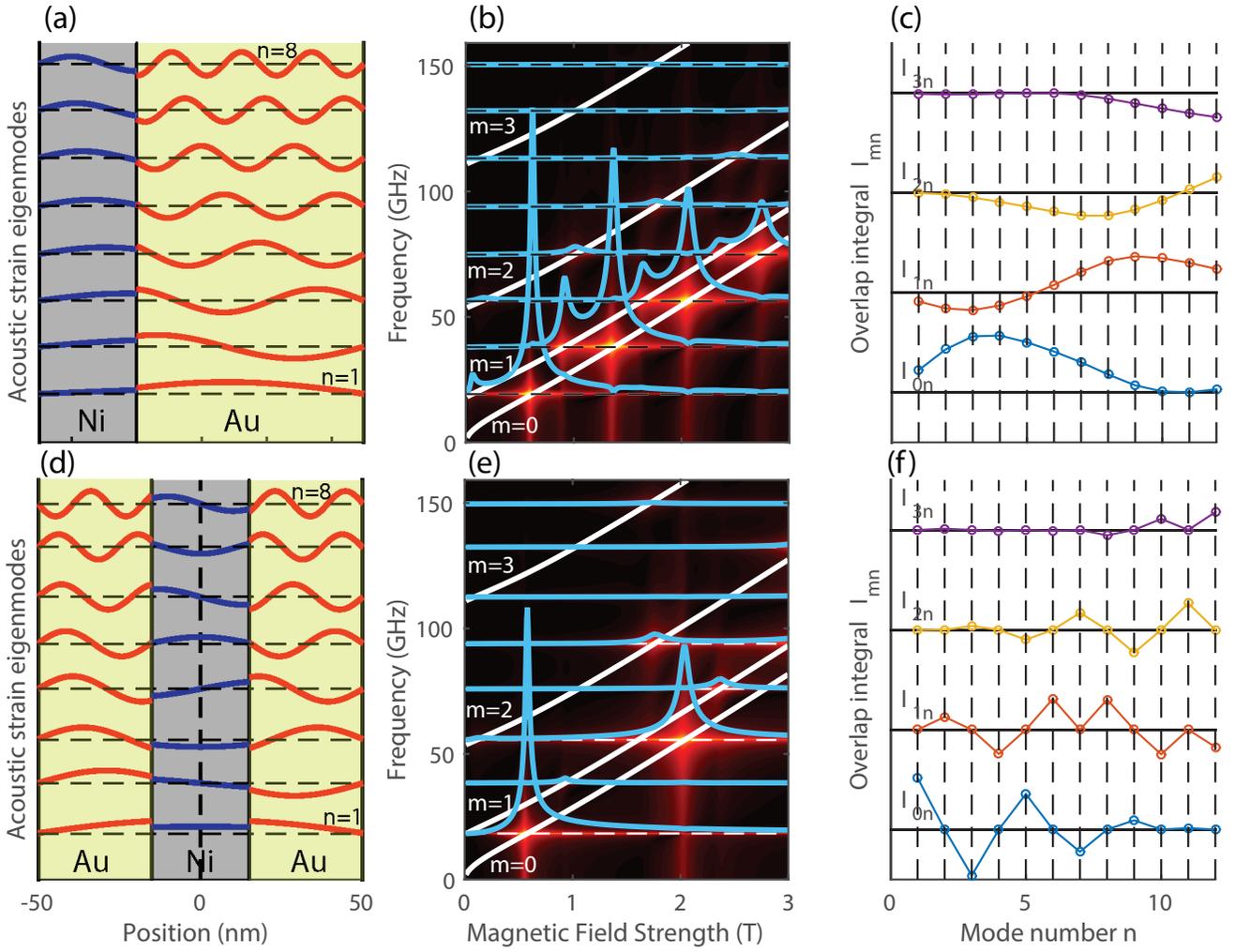}
 \caption{\label{Fig4_new} (a) Profiles of the first eight acoustic strain eigenmodes. (b) The magnetic field dependence of the elastically induced magnetization dynamics according to Eq.~(\ref{m_H_omega}). (c) The overlap integral $I_{mn}$ for the first four magnon modes.}
\end{figure*}

In order to illustrate the peculiarities of phonon-magnon interactions we decided to compare two different freestanding multilayer structures: a 30~nm Ni/70~nm Au bilayer and a symmetric 35~nm Au/30~nm Ni/35~nm Au trilayer of the same total thickness of 100~nm.  The results for both structures are presented in Fig.~4(a-c) and Fig.~4(d-f), respectively. Whereas the acoustic spectra $\omega_n$ for both spectra are nearly identical, their eigenmodes are fundamentally different. All eigenmodes for the bilayer (Fig.~4(a)) are naturally asymmetric. In contrast, eigenmodes in the trilayer in Fig.4~(d) split in two sub-families: symmetric (for $n=1,3,5,...$) and antisymmetric (for $n=2,4,6,...$) modes with respect to the symmetry plane in the center of the Ni-layer. This intrinsic difference in the symmetry between the two cases has important consequences for phonon-magnon interactions given that magnon modes have their own and well defined symmetry and are also split in two sub-families: symmetric (for $m=0,2,4,...$) and antisymmetric (for $m=1,3,5,...$) . Panels (b) and (e) compare the Fourier spectra of magnetization dynamics obtained with Eq.~(\ref{m_H_omega}), generalized with acoustic damping.
Figure 4(b) clearly demonstrates that the resonant enhancement of magnetic signal when magnon dispersions (white continuous lines) cross acoustic resonances (blue dashed lines). This indicates that prefactors $P_m(H)a_nI_{nm}$ are nonzero. Moreover, due to the non-zero Gilbert damping, individual magnetic resonances excited by the same acoustic mode, overlap: see the continuous blue profiles in Fig.~4(b). 
The situation becomes fundamentally different for the trilayer in Fig.~4(e): only a few resonances are excited and they do not overlap. This indicates that some of the prefactors $P_m(H)a_nI_{nm}$ must be zero. This can be understood based on the symmetry considerations reported above. Indeed, since in this case both modes, magnonic and acoustic, have well defined symmetries split into two families. Thereby, as a detailed analysis confirms, when the magnonic and acoustic modes belong to different symmetry sub-families the overlap integral $I_{mn}$ becomes negligibly small. As a consequence, whereas for the bilayer the $I_{mn}$ is nonzero for most of the modes (Fig.~4(c)), for the trilayer it displays many zeros (Fig.~4(f)).  Applied to the plot in Fig.~4(e), these symmetry considerations result in that the symmmetric $m=0,2,..$ (antisymmertic $m=1,3,..$) magnon modes should only interact with symmetric $n=1,3,...$ (antisymmetric $n=2,4,...$) acoustic modes. In other words the most closely spaced $m=0$ and $m=1$ magnon modes will never interact with the same acoustic modes and their resonances will not overlap. As such, the freestanding metal/ferromagnet/metal trilayers structures can serve as an excellent playground to investigate individual phonon-magnon interactions.

\begin{figure*} 
	\footnotesize{} \centering
 \includegraphics[width=2.0\columnwidth]{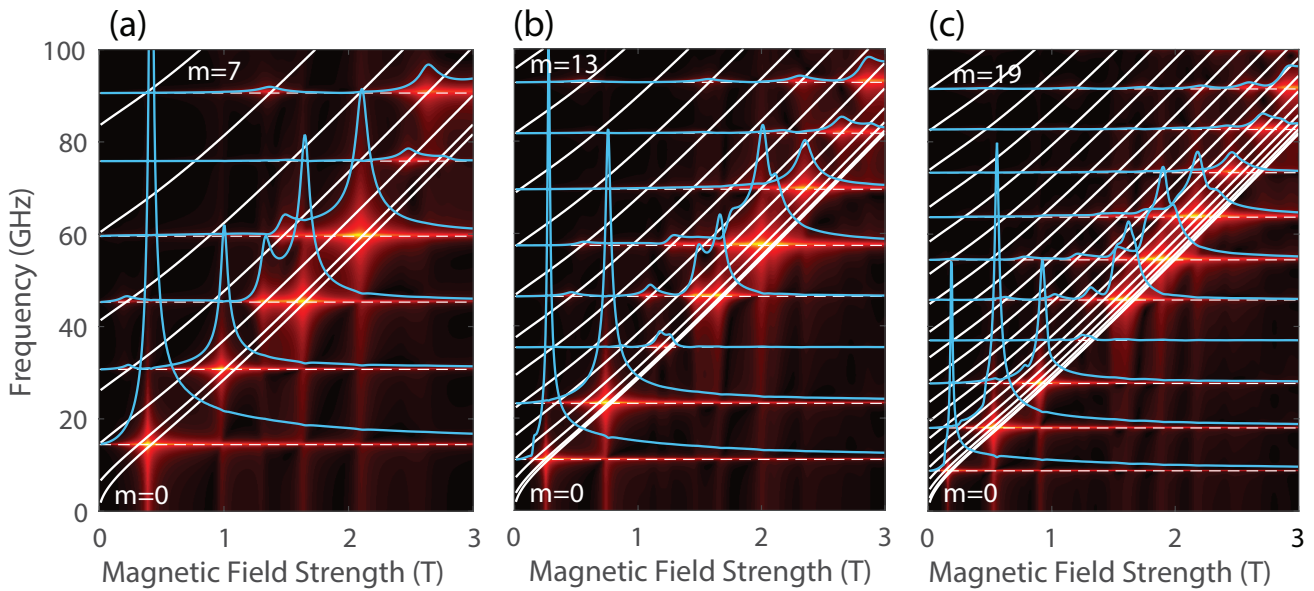}
 \caption{\label{fig5_new} The magnetic field dependence of the elastically induced magnetization dynamics for three symmetric Au(35~nm)/Ni (X~nm)/Au(35~nm)-trilayer structures: (a) $2l_1$=70~nm, (b) $2l_1$=130~nm and (c) $2l_1$=200~nm.}
\end{figure*}

In order to illustrate the capability of symmetric freestanding trilayers to separate individual phonon-magnon resonances, we have performed the very same simulation for a series of structures with varying Ni thickness and the same thickness of the gold layer of 35~nm. Figure 5 compares the magnetization response for Ni thickness of 70, 130 and 200~nm, respectively. With increasing Ni thickness the spectral density of acoustic and magnetic modes increases. Nevertheless the magneto-elastic resonances driven by the two lowest order acoustic modes $n$=1 and $n$=2 remain clearly resolved for all structures. The situation becomes very different for $n$=3-driven resonances: for $2l_1$=70~nm is excites partially overlapping $m$=0 and $m$=2 magnons with a significant amplitude. For $2l_1$=130~nm this resonance almost disappears and for $2l_1$=200~nm becomes larger again. In the latter case, however, the $n$=4-driven resonance disappears. For higher $n$ resonances with multiple $m$ overlap giving rise to broad complex-shaped peaks of large amplitude. It should be noticed that the $n$=2-driven magneto-elastic resonance remains well-defined and its amplitude is growing with the nickel thickness, becoming larger that $n$=1-driven FMR-resonance.
The above qualitative analysis shows that metal/ferromagnet/metal trilayers allow allow for an unambiguous discrimination between magneto-elastic excitation of ($m=0$) FMR and ($m\leq 1$) exchange magnon modes due to the symmetry of acoustic eigenmodes, for a wide range of structures with the thickness from 100 to 270~nm. 

This situation corresponds very well to the capabilities to produce macroscopically large samples of suspended membranes: either using chemical means \cite{kim2017magnetization} or nondestructive nanostructuring of thin films using single ultrashort laser pulses \cite{domke2014understanding,temnov2020nondestructive}.  

It appears to be instructive, at least from the theoretical perspective, to analyse phonon-magnon interactions for ultrathin multilayers. Figure 6 displays the results for a Au(3~nm)/Ni(10~nm)/Au(3~nm) trilayer. It is remarkable that magnons with frequencies of several hundreds GHz be excited: see  $m=2,n=3$ and $m$=2,$n$=5 resonances at 434~GHz and 707~GHz, respectively.  Moreover, the  $m$=2,$n$=3 magneto-elastic resonance at 434~GHz should be observable at relatively low, experimentally easily accessible external magnetic fields with  amplitude of approximately 0.8~Tesla.

\begin{figure*} 
	\footnotesize{} \centering
 \includegraphics[width=2.0\columnwidth]{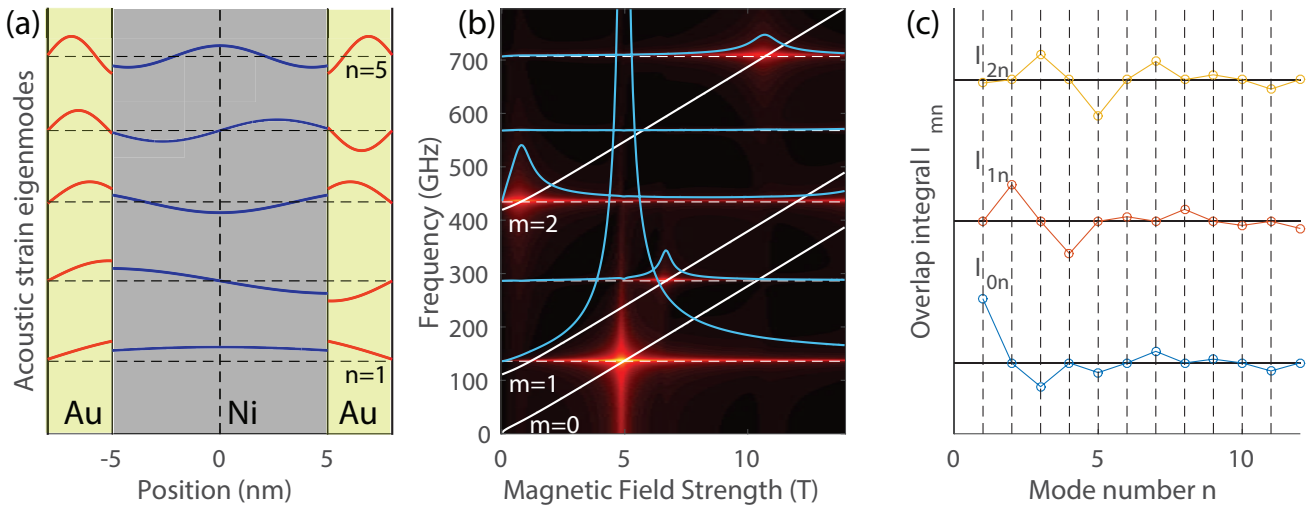}
 \caption{\label{Fig6_new} The magnetic field dependence of the elastically induced magnetization dynamics for a symmetric Au(3~nm)/Ni(10~nm)/Au(3~nm)-trilayer structure demonstrates the possibility of magneto-elastic excitation of exchange magnons in the Thz-frequency range.}
\end{figure*}

\section{Discussion}
The theoretical analysis in this paper can be particularly useful in the context of recent magneto-elastic studies in freestanding ferromagnet/dielectric/ferromagnet trilayers discussing the strong acoustically mediated coupling between FMR excitations in two identical ferromagnetic (YIG) layers \cite{an2020coherent,an2022bright}. To be more specific, it has been shown that the magneto-elastic coupling between FMR-oscillators in a YIG1(200~nm)|GGG(0.5~mm)|YIG2(200~nm) results in the formation of the bandgap \cite{an2020coherent} as well as bright and dark eigenmodes of two coupled oscillators \cite{an2022bright}. It would be very interesting to investigate if the same mechanism could work for exchange magnons with $n\geq 1$ in YIG layers. The detailed understanding of acoustic modes in the present paper could be quite helpful.

Another recent work \cite{jaris2020towards,yang2021acoustic} monitors the excitation of disc nanomagnets with diameter $D$ using monochromatic surface acoustic waves with wavelength $\lambda_{SAW}$. It reports on the increase of the effective Gilbert damping in the regime $D>\lambda_{SAW}$, which is associated to the effect of inhomogeneous broadening in ensemble measurements. At the same time, it follows from our analysis that magneto-elastic excitation of (in-plane) magnons can modify the shape of magnetic response though overlapping of FMR and magnon resonances. Spatially inhomogeneous magnon modes could contribute to the time-resolved MOKE-microscopy signals due to the enhanced sensitivity reported on discs of similar dimensions \cite{berger2015enhanced}.       

From the theoretical perspective the simplified description of magnon excitation based on Eq.~(\ref{magnon_oscillator}) can be generalized model the time-dependent solutions in ultrafast magneto-elastic interactions \cite{BesseJMMM,jaris2020towards,yang2021acoustic}. Among different non-stationary processes the magneto-elastic parametric effects \cite{chang2017parametric} play the particular role since they are sensitive both to the Gilbert damping and the amplitude of the parametric modulation. 

\section{Summary and conclusions}
The main result of this paper is the relatively simple approach to describe ultrafast (GHz-to-THz frequency) magneto-acoustic interactions in free-standing ferromagnetic multilayers on the basis of eigenmode decomposition, both for for magnons and phonons. Semi-analytical calculations of acoustic eigenmodes in freestanding bilayers are generalized to describe symmetric trilayer structures. The numerically calculated acoustic eigenmodes are used to quantify resonantly enhanced interactions between individual phonon and magnon modes. Whereas in asymmetric bilayers nearly all phonon and magnon modes are coupled, the symmetry of trilayers results in the suppression of certain magnon-phonon interactions due to the zero overlap integrals of eigenmodes. This fact should allow for unambiguous discrimination between phonon-($m=0$)FMR and phonon-($m\geq 1$)magnon interactions in ultrafast magneto-acoustic experiments.  Several forward-looking statements have been made with respect to the existing experimental configurations.               


\begin{acknowledgments}
Financial support by the ANR-21-CE15-0048-01 MRSEI "IRON-MAG" is gratefully acknowledged.
\end{acknowledgments}


\begin{thebibliography}{25}%
\makeatletter
\providecommand \@ifxundefined [1]{%
 \@ifx{#1\undefined}
}%
\providecommand \@ifnum [1]{%
 \ifnum #1\expandafter \@firstoftwo
 \else \expandafter \@secondoftwo
 \fi
}%
\providecommand \@ifx [1]{%
 \ifx #1\expandafter \@firstoftwo
 \else \expandafter \@secondoftwo
 \fi
}%
\providecommand \natexlab [1]{#1}%
\providecommand \enquote  [1]{``#1''}%
\providecommand \bibnamefont  [1]{#1}%
\providecommand \bibfnamefont [1]{#1}%
\providecommand \citenamefont [1]{#1}%
\providecommand \href@noop [0]{\@secondoftwo}%
\providecommand \href [0]{\begingroup \@sanitize@url \@href}%
\providecommand \@href[1]{\@@startlink{#1}\@@href}%
\providecommand \@@href[1]{\endgroup#1\@@endlink}%
\providecommand \@sanitize@url [0]{\catcode `\\12\catcode `\$12\catcode
  `\&12\catcode `\#12\catcode `\^12\catcode `\_12\catcode `\%12\relax}%
\providecommand \@@startlink[1]{}%
\providecommand \@@endlink[0]{}%
\providecommand \url  [0]{\begingroup\@sanitize@url \@url }%
\providecommand \@url [1]{\endgroup\@href {#1}{\urlprefix }}%
\providecommand \urlprefix  [0]{URL }%
\providecommand \Eprint [0]{\href }%
\providecommand \doibase [0]{http://dx.doi.org/}%
\providecommand \selectlanguage [0]{\@gobble}%
\providecommand \bibinfo  [0]{\@secondoftwo}%
\providecommand \bibfield  [0]{\@secondoftwo}%
\providecommand \translation [1]{[#1]}%
\providecommand \BibitemOpen [0]{}%
\providecommand \bibitemStop [0]{}%
\providecommand \bibitemNoStop [0]{.\EOS\space}%
\providecommand \EOS [0]{\spacefactor3000\relax}%
\providecommand \BibitemShut  [1]{\csname bibitem#1\endcsname}%
\let\auto@bib@innerbib\@empty
\bibitem [{\citenamefont {Scherbakov}\ \emph {et~al.}(2010)\citenamefont
  {Scherbakov}, \citenamefont {Salasyuk}, \citenamefont {Akimov}, \citenamefont
  {Liu}, \citenamefont {Bombeck}, \citenamefont {Br\"uggemann}, \citenamefont
  {Yakovlev}, \citenamefont {Sapega}, \citenamefont {Furdyna},\ and\
  \citenamefont {Bayer}}]{Scherbakov2010}%
  \BibitemOpen
  \bibfield  {author} {\bibinfo {author} {\bibfnamefont {A.~V.}\ \bibnamefont
  {Scherbakov}}, \bibinfo {author} {\bibfnamefont {A.~S.}\ \bibnamefont
  {Salasyuk}}, \bibinfo {author} {\bibfnamefont {A.~V.}\ \bibnamefont
  {Akimov}}, \bibinfo {author} {\bibfnamefont {X.}~\bibnamefont {Liu}},
  \bibinfo {author} {\bibfnamefont {M.}~\bibnamefont {Bombeck}}, \bibinfo
  {author} {\bibfnamefont {C.}~\bibnamefont {Br\"uggemann}}, \bibinfo {author}
  {\bibfnamefont {D.~R.}\ \bibnamefont {Yakovlev}}, \bibinfo {author}
  {\bibfnamefont {V.~F.}\ \bibnamefont {Sapega}}, \bibinfo {author}
  {\bibfnamefont {J.~K.}\ \bibnamefont {Furdyna}}, \ and\ \bibinfo {author}
  {\bibfnamefont {M.}~\bibnamefont {Bayer}},\ }\href {\doibase
  10.1103/PhysRevLett.105.117204} {\bibfield  {journal} {\bibinfo  {journal}
  {Phys. Rev. Lett.}\ }\textbf {\bibinfo {volume} {105}},\ \bibinfo {pages}
  {117204} (\bibinfo {year} {2010})}\BibitemShut {NoStop}%
\bibitem [{\citenamefont {Kim}\ \emph {et~al.}(2012)\citenamefont {Kim},
  \citenamefont {Vomir},\ and\ \citenamefont {Bigot}}]{kim2012ultrafast}%
  \BibitemOpen
  \bibfield  {author} {\bibinfo {author} {\bibfnamefont {J.-W.}\ \bibnamefont
  {Kim}}, \bibinfo {author} {\bibfnamefont {M.}~\bibnamefont {Vomir}}, \ and\
  \bibinfo {author} {\bibfnamefont {J.-Y.}\ \bibnamefont {Bigot}},\ }\href
  {\doibase 10.1103/PhysRevLett.109.166601} {\bibfield  {journal} {\bibinfo
  {journal} {Phys. Rev. Lett.}\ }\textbf {\bibinfo {volume} {109}},\ \bibinfo
  {pages} {166601} (\bibinfo {year} {2012})}\BibitemShut {NoStop}%
\bibitem [{\citenamefont {Thevenard}\ \emph {et~al.}(2010)\citenamefont
  {Thevenard}, \citenamefont {Peronne}, \citenamefont {Gourdon}, \citenamefont
  {Testelin}, \citenamefont {Cubukcu}, \citenamefont {Charron}, \citenamefont
  {Vincent}, \citenamefont {Lema\^{\i}tre},\ and\ \citenamefont
  {Perrin}}]{Thevenard2010}%
  \BibitemOpen
  \bibfield  {author} {\bibinfo {author} {\bibfnamefont {L.}~\bibnamefont
  {Thevenard}}, \bibinfo {author} {\bibfnamefont {E.}~\bibnamefont {Peronne}},
  \bibinfo {author} {\bibfnamefont {C.}~\bibnamefont {Gourdon}}, \bibinfo
  {author} {\bibfnamefont {C.}~\bibnamefont {Testelin}}, \bibinfo {author}
  {\bibfnamefont {M.}~\bibnamefont {Cubukcu}}, \bibinfo {author} {\bibfnamefont
  {E.}~\bibnamefont {Charron}}, \bibinfo {author} {\bibfnamefont
  {S.}~\bibnamefont {Vincent}}, \bibinfo {author} {\bibfnamefont
  {A.}~\bibnamefont {Lema\^{\i}tre}}, \ and\ \bibinfo {author} {\bibfnamefont
  {B.}~\bibnamefont {Perrin}},\ }\href {\doibase 10.1103/PhysRevB.82.104422}
  {\bibfield  {journal} {\bibinfo  {journal} {Phys. Rev. B}\ }\textbf {\bibinfo
  {volume} {82}},\ \bibinfo {pages} {104422} (\bibinfo {year}
  {2010})}\BibitemShut {NoStop}%
\bibitem [{\citenamefont {Janu{\v{s}}onis}\ \emph {et~al.}(2016)\citenamefont
  {Janu{\v{s}}onis}, \citenamefont {Chang}, \citenamefont {Jansma},
  \citenamefont {Gatilova}, \citenamefont {Vlasov}, \citenamefont {Lomonosov},
  \citenamefont {Temnov},\ and\ \citenamefont
  {Tobey}}]{janusonis2016ultrafast}%
  \BibitemOpen
  \bibfield  {author} {\bibinfo {author} {\bibfnamefont {J.}~\bibnamefont
  {Janu{\v{s}}onis}}, \bibinfo {author} {\bibfnamefont {C.~L.}\ \bibnamefont
  {Chang}}, \bibinfo {author} {\bibfnamefont {T.}~\bibnamefont {Jansma}},
  \bibinfo {author} {\bibfnamefont {A.}~\bibnamefont {Gatilova}}, \bibinfo
  {author} {\bibfnamefont {V.~S.}\ \bibnamefont {Vlasov}}, \bibinfo {author}
  {\bibfnamefont {A.~M.}\ \bibnamefont {Lomonosov}}, \bibinfo {author}
  {\bibfnamefont {V.~V.}\ \bibnamefont {Temnov}}, \ and\ \bibinfo {author}
  {\bibfnamefont {R.~I.}\ \bibnamefont {Tobey}},\ }\href@noop {} {\bibfield
  {journal} {\bibinfo  {journal} {Phys. Rev. B}\ }\textbf {\bibinfo {volume}
  {94}},\ \bibinfo {pages} {024415} (\bibinfo {year} {2016})}\BibitemShut
  {NoStop}%
\bibitem [{\citenamefont {Chang}\ \emph {et~al.}(2017)\citenamefont {Chang},
  \citenamefont {Lomonosov}, \citenamefont {Janusonis}, \citenamefont {Vlasov},
  \citenamefont {Temnov},\ and\ \citenamefont {Tobey}}]{chang2017parametric}%
  \BibitemOpen
  \bibfield  {author} {\bibinfo {author} {\bibfnamefont {C.~L.}\ \bibnamefont
  {Chang}}, \bibinfo {author} {\bibfnamefont {A.~M.}\ \bibnamefont
  {Lomonosov}}, \bibinfo {author} {\bibfnamefont {J.}~\bibnamefont
  {Janusonis}}, \bibinfo {author} {\bibfnamefont {V.~S.}\ \bibnamefont
  {Vlasov}}, \bibinfo {author} {\bibfnamefont {V.~V.}\ \bibnamefont {Temnov}},
  \ and\ \bibinfo {author} {\bibfnamefont {R.~I.}\ \bibnamefont {Tobey}},\
  }\href@noop {} {\bibfield  {journal} {\bibinfo  {journal} {Phys. Rev. B}\
  }\textbf {\bibinfo {volume} {95}},\ \bibinfo {pages} {060409(R)} (\bibinfo
  {year} {2017})}\BibitemShut {NoStop}%
\bibitem [{\citenamefont {Vlasov}\ \emph {et~al.}(2022)\citenamefont {Vlasov},
  \citenamefont {Golov}, \citenamefont {Kotov}, \citenamefont {Shcheglov},
  \citenamefont {Lomonosov},\ and\ \citenamefont {Temnov}}]{Vlasov_2022}%
  \BibitemOpen
  \bibfield  {author} {\bibinfo {author} {\bibfnamefont {V.~S.}\ \bibnamefont
  {Vlasov}}, \bibinfo {author} {\bibfnamefont {A.~V.}\ \bibnamefont {Golov}},
  \bibinfo {author} {\bibfnamefont {L.~N.}\ \bibnamefont {Kotov}}, \bibinfo
  {author} {\bibfnamefont {V.~I.}\ \bibnamefont {Shcheglov}}, \bibinfo {author}
  {\bibfnamefont {A.~M.}\ \bibnamefont {Lomonosov}}, \ and\ \bibinfo {author}
  {\bibfnamefont {V.~V.}\ \bibnamefont {Temnov}},\ }\href@noop {} {\bibfield
  {journal} {\bibinfo  {journal} {Acoustical Physics}\ }\textbf {\bibinfo
  {volume} {68}},\ \bibinfo {pages} {18} (\bibinfo {year} {2022})}\BibitemShut
  {NoStop}%
\bibitem [{\citenamefont {Thevenard}\ \emph {et~al.}(2013)\citenamefont
  {Thevenard}, \citenamefont {Duquesne}, \citenamefont {Peronne}, \citenamefont
  {Von~Bardeleben}, \citenamefont {Jaffres}, \citenamefont {Ruttala},
  \citenamefont {George}, \citenamefont {Lemaitre},\ and\ \citenamefont
  {Gourdon}}]{thevenard2013irreversible}%
  \BibitemOpen
  \bibfield  {author} {\bibinfo {author} {\bibfnamefont {L.}~\bibnamefont
  {Thevenard}}, \bibinfo {author} {\bibfnamefont {J.-Y.}\ \bibnamefont
  {Duquesne}}, \bibinfo {author} {\bibfnamefont {E.}~\bibnamefont {Peronne}},
  \bibinfo {author} {\bibfnamefont {H.~J.}\ \bibnamefont {Von~Bardeleben}},
  \bibinfo {author} {\bibfnamefont {H.}~\bibnamefont {Jaffres}}, \bibinfo
  {author} {\bibfnamefont {S.}~\bibnamefont {Ruttala}}, \bibinfo {author}
  {\bibfnamefont {J.-M.}\ \bibnamefont {George}}, \bibinfo {author}
  {\bibfnamefont {A.}~\bibnamefont {Lemaitre}}, \ and\ \bibinfo {author}
  {\bibfnamefont {C.}~\bibnamefont {Gourdon}},\ }\href@noop {} {\bibfield
  {journal} {\bibinfo  {journal} {Physical Review B}\ }\textbf {\bibinfo
  {volume} {87}},\ \bibinfo {pages} {144402} (\bibinfo {year}
  {2013})}\BibitemShut {NoStop}%
\bibitem [{\citenamefont {Vlasov}\ \emph {et~al.}(2020)\citenamefont {Vlasov},
  \citenamefont {Lomonosov}, \citenamefont {Golov}, \citenamefont {Kotov},
  \citenamefont {Besse}, \citenamefont {Alekhin}, \citenamefont {Kuzmin},
  \citenamefont {Bychkov},\ and\ \citenamefont {Temnov}}]{VlasovPRB2020}%
  \BibitemOpen
  \bibfield  {author} {\bibinfo {author} {\bibfnamefont {V.~S.}\ \bibnamefont
  {Vlasov}}, \bibinfo {author} {\bibfnamefont {A.~M.}\ \bibnamefont
  {Lomonosov}}, \bibinfo {author} {\bibfnamefont {A.~V.}\ \bibnamefont
  {Golov}}, \bibinfo {author} {\bibfnamefont {L.~N.}\ \bibnamefont {Kotov}},
  \bibinfo {author} {\bibfnamefont {V.}~\bibnamefont {Besse}}, \bibinfo
  {author} {\bibfnamefont {A.}~\bibnamefont {Alekhin}}, \bibinfo {author}
  {\bibfnamefont {D.~A.}\ \bibnamefont {Kuzmin}}, \bibinfo {author}
  {\bibfnamefont {I.~V.}\ \bibnamefont {Bychkov}}, \ and\ \bibinfo {author}
  {\bibfnamefont {V.~V.}\ \bibnamefont {Temnov}},\ }\href {\doibase
  10.1103/PhysRevB.101.024425} {\bibfield  {journal} {\bibinfo  {journal}
  {Phys. Rev. B}\ }\textbf {\bibinfo {volume} {101}},\ \bibinfo {pages}
  {024425} (\bibinfo {year} {2020})}\BibitemShut {NoStop}%
\bibitem [{\citenamefont {Berk}\ \emph {et~al.}(2019)\citenamefont {Berk},
  \citenamefont {Jaris}, \citenamefont {Yang}, \citenamefont {Dhuey},
  \citenamefont {Cabrini},\ and\ \citenamefont {Schmidt}}]{berk2019strongly}%
  \BibitemOpen
  \bibfield  {author} {\bibinfo {author} {\bibfnamefont {C.}~\bibnamefont
  {Berk}}, \bibinfo {author} {\bibfnamefont {M.}~\bibnamefont {Jaris}},
  \bibinfo {author} {\bibfnamefont {W.}~\bibnamefont {Yang}}, \bibinfo {author}
  {\bibfnamefont {S.}~\bibnamefont {Dhuey}}, \bibinfo {author} {\bibfnamefont
  {S.}~\bibnamefont {Cabrini}}, \ and\ \bibinfo {author} {\bibfnamefont
  {H.}~\bibnamefont {Schmidt}},\ }\href@noop {} {\bibfield  {journal} {\bibinfo
   {journal} {Nature communications}\ }\textbf {\bibinfo {volume} {10}},\
  \bibinfo {pages} {1} (\bibinfo {year} {2019})}\BibitemShut {NoStop}%
\bibitem [{\citenamefont {Dianov}\ \emph {et~al.}(2013)\citenamefont {Dianov},
  \citenamefont {Vlasov}, \citenamefont {Beznosikov}, \citenamefont {Kotov},
  \citenamefont {Shavrov},\ and\ \citenamefont {Shcheglov}}]{Vlasov_2013}%
  \BibitemOpen
  \bibfield  {author} {\bibinfo {author} {\bibfnamefont {M.~Y.}\ \bibnamefont
  {Dianov}}, \bibinfo {author} {\bibfnamefont {V.~S.}\ \bibnamefont {Vlasov}},
  \bibinfo {author} {\bibfnamefont {D.~S.}\ \bibnamefont {Beznosikov}},
  \bibinfo {author} {\bibfnamefont {L.~N.}\ \bibnamefont {Kotov}}, \bibinfo
  {author} {\bibfnamefont {V.~G.}\ \bibnamefont {Shavrov}}, \ and\ \bibinfo
  {author} {\bibfnamefont {V.~I.}\ \bibnamefont {Shcheglov}},\ }\href@noop {}
  {\bibfield  {journal} {\bibinfo  {journal} {Bulletin of the Russian Academy
  of Sciences. Physics}\ }\textbf {\bibinfo {volume} {77}},\ \bibinfo {pages}
  {1258} (\bibinfo {year} {2013})}\BibitemShut {NoStop}%
\bibitem [{\citenamefont {Vlasov}\ \emph {et~al.}(2014)\citenamefont {Vlasov},
  \citenamefont {Shavrov},\ and\ \citenamefont {Shcheglov}}]{Vlasov_2014}%
  \BibitemOpen
  \bibfield  {author} {\bibinfo {author} {\bibfnamefont {V.~S.}\ \bibnamefont
  {Vlasov}}, \bibinfo {author} {\bibfnamefont {V.~G.}\ \bibnamefont {Shavrov}},
  \ and\ \bibinfo {author} {\bibfnamefont {V.~I.}\ \bibnamefont {Shcheglov}},\
  }\href@noop {} {\bibfield  {journal} {\bibinfo  {journal} {Journal of
  Communications Technology and Electronics}\ }\textbf {\bibinfo {volume}
  {59}},\ \bibinfo {pages} {441} (\bibinfo {year} {2014})}\BibitemShut
  {NoStop}%
\bibitem [{\citenamefont {Van~Kampen}\ \emph {et~al.}(2002)\citenamefont
  {Van~Kampen}, \citenamefont {Jozsa}, \citenamefont {Kohlhepp}, \citenamefont
  {LeClair}, \citenamefont {Lagae}, \citenamefont {De~Jonge},\ and\
  \citenamefont {Koopmans}}]{van2002all}%
  \BibitemOpen
  \bibfield  {author} {\bibinfo {author} {\bibfnamefont {M.}~\bibnamefont
  {Van~Kampen}}, \bibinfo {author} {\bibfnamefont {C.}~\bibnamefont {Jozsa}},
  \bibinfo {author} {\bibfnamefont {J.}~\bibnamefont {Kohlhepp}}, \bibinfo
  {author} {\bibfnamefont {P.}~\bibnamefont {LeClair}}, \bibinfo {author}
  {\bibfnamefont {L.}~\bibnamefont {Lagae}}, \bibinfo {author} {\bibfnamefont
  {W.}~\bibnamefont {De~Jonge}}, \ and\ \bibinfo {author} {\bibfnamefont
  {B.}~\bibnamefont {Koopmans}},\ }\href {\doibase
  10.1103/PhysRevLett.88.227201} {\bibfield  {journal} {\bibinfo  {journal}
  {Phys. Rev. Lett.}\ }\textbf {\bibinfo {volume} {88}},\ \bibinfo {pages}
  {227201} (\bibinfo {year} {2002})}\BibitemShut {NoStop}%
\bibitem [{\citenamefont {Bombeck}\ \emph {et~al.}(2012)\citenamefont
  {Bombeck}, \citenamefont {Salasyuk}, \citenamefont {Glavin}, \citenamefont
  {Scherbakov}, \citenamefont {Br{\"u}ggemann}, \citenamefont {Yakovlev},
  \citenamefont {Sapega}, \citenamefont {Liu}, \citenamefont {Furdyna},
  \citenamefont {Akimov} \emph {et~al.}}]{bombeck2012excitation}%
  \BibitemOpen
  \bibfield  {author} {\bibinfo {author} {\bibfnamefont {M.}~\bibnamefont
  {Bombeck}}, \bibinfo {author} {\bibfnamefont {A.}~\bibnamefont {Salasyuk}},
  \bibinfo {author} {\bibfnamefont {B.}~\bibnamefont {Glavin}}, \bibinfo
  {author} {\bibfnamefont {A.}~\bibnamefont {Scherbakov}}, \bibinfo {author}
  {\bibfnamefont {C.}~\bibnamefont {Br{\"u}ggemann}}, \bibinfo {author}
  {\bibfnamefont {D.}~\bibnamefont {Yakovlev}}, \bibinfo {author}
  {\bibfnamefont {V.}~\bibnamefont {Sapega}}, \bibinfo {author} {\bibfnamefont
  {X.}~\bibnamefont {Liu}}, \bibinfo {author} {\bibfnamefont {J.}~\bibnamefont
  {Furdyna}}, \bibinfo {author} {\bibfnamefont {A.}~\bibnamefont {Akimov}},
  \emph {et~al.},\ }\href {\doibase 10.1103/PhysRevB.85.195324} {\bibfield
  {journal} {\bibinfo  {journal} {Phys. Rev. B}\ }\textbf {\bibinfo {volume}
  {85}},\ \bibinfo {pages} {195324} (\bibinfo {year} {2012})}\BibitemShut
  {NoStop}%
\bibitem [{\citenamefont {Besse}\ \emph {et~al.}(2020)\citenamefont {Besse},
  \citenamefont {Golov}, \citenamefont {Vlasov}, \citenamefont {Alekhin},
  \citenamefont {Kuzmin}, \citenamefont {Bychkov}, \citenamefont {Kotov},\ and\
  \citenamefont {Temnov}}]{BesseJMMM}%
  \BibitemOpen
  \bibfield  {author} {\bibinfo {author} {\bibfnamefont {V.}~\bibnamefont
  {Besse}}, \bibinfo {author} {\bibfnamefont {A.~V.}\ \bibnamefont {Golov}},
  \bibinfo {author} {\bibfnamefont {V.~S.}\ \bibnamefont {Vlasov}}, \bibinfo
  {author} {\bibfnamefont {A.}~\bibnamefont {Alekhin}}, \bibinfo {author}
  {\bibfnamefont {D.}~\bibnamefont {Kuzmin}}, \bibinfo {author} {\bibfnamefont
  {I.~V.}\ \bibnamefont {Bychkov}}, \bibinfo {author} {\bibfnamefont {L.~N.}\
  \bibnamefont {Kotov}}, \ and\ \bibinfo {author} {\bibfnamefont {V.~V.}\
  \bibnamefont {Temnov}},\ }\href@noop {} {\bibfield  {journal} {\bibinfo
  {journal} {J. Magn. Magn. Mater.}\ }\textbf {\bibinfo {volume} {502}},\
  \bibinfo {pages} {166320} (\bibinfo {year} {2020})}\BibitemShut {NoStop}%
\bibitem [{\citenamefont {Kim}\ and\ \citenamefont
  {Bigot}(2017)}]{kim2017magnetization}%
  \BibitemOpen
  \bibfield  {author} {\bibinfo {author} {\bibfnamefont {J.-W.}\ \bibnamefont
  {Kim}}\ and\ \bibinfo {author} {\bibfnamefont {J.-Y.}\ \bibnamefont
  {Bigot}},\ }\href {\doibase 10.1103/PhysRevB.95.144422} {\bibfield  {journal}
  {\bibinfo  {journal} {Phys. Rev. B}\ }\textbf {\bibinfo {volume} {95}},\
  \bibinfo {pages} {144422} (\bibinfo {year} {2017})}\BibitemShut {NoStop}%
\bibitem [{\citenamefont {An}\ \emph {et~al.}(2020)\citenamefont {An},
  \citenamefont {Litvinenko}, \citenamefont {Kohno}, \citenamefont {Fuad},
  \citenamefont {Naletov}, \citenamefont {Vila}, \citenamefont {Ebels},
  \citenamefont {de~Loubens}, \citenamefont {Hurdequint}, \citenamefont
  {Beaulieu} \emph {et~al.}}]{an2020coherent}%
  \BibitemOpen
  \bibfield  {author} {\bibinfo {author} {\bibfnamefont {K.}~\bibnamefont
  {An}}, \bibinfo {author} {\bibfnamefont {A.~N.}\ \bibnamefont {Litvinenko}},
  \bibinfo {author} {\bibfnamefont {R.}~\bibnamefont {Kohno}}, \bibinfo
  {author} {\bibfnamefont {A.~A.}\ \bibnamefont {Fuad}}, \bibinfo {author}
  {\bibfnamefont {V.~V.}\ \bibnamefont {Naletov}}, \bibinfo {author}
  {\bibfnamefont {L.}~\bibnamefont {Vila}}, \bibinfo {author} {\bibfnamefont
  {U.}~\bibnamefont {Ebels}}, \bibinfo {author} {\bibfnamefont
  {G.}~\bibnamefont {de~Loubens}}, \bibinfo {author} {\bibfnamefont
  {H.}~\bibnamefont {Hurdequint}}, \bibinfo {author} {\bibfnamefont
  {N.}~\bibnamefont {Beaulieu}},  \emph {et~al.},\ }\href@noop {} {\bibfield
  {journal} {\bibinfo  {journal} {Physical Review B}\ }\textbf {\bibinfo
  {volume} {101}},\ \bibinfo {pages} {060407} (\bibinfo {year}
  {2020})}\BibitemShut {NoStop}%
\bibitem [{\citenamefont {An}\ \emph {et~al.}(2022)\citenamefont {An},
  \citenamefont {Kohno}, \citenamefont {Litvinenko}, \citenamefont {Seeger},
  \citenamefont {Naletov}, \citenamefont {Vila}, \citenamefont {de~Loubens},
  \citenamefont {Youssef}, \citenamefont {Vukadinovic}, \citenamefont {Bauer}
  \emph {et~al.}}]{an2022bright}%
  \BibitemOpen
  \bibfield  {author} {\bibinfo {author} {\bibfnamefont {K.}~\bibnamefont
  {An}}, \bibinfo {author} {\bibfnamefont {R.}~\bibnamefont {Kohno}}, \bibinfo
  {author} {\bibfnamefont {A.~N.}\ \bibnamefont {Litvinenko}}, \bibinfo
  {author} {\bibfnamefont {R.~L.}\ \bibnamefont {Seeger}}, \bibinfo {author}
  {\bibfnamefont {V.~V.}\ \bibnamefont {Naletov}}, \bibinfo {author}
  {\bibfnamefont {L.}~\bibnamefont {Vila}}, \bibinfo {author} {\bibfnamefont
  {G.}~\bibnamefont {de~Loubens}}, \bibinfo {author} {\bibfnamefont {J.~B.}\
  \bibnamefont {Youssef}}, \bibinfo {author} {\bibfnamefont {N.}~\bibnamefont
  {Vukadinovic}}, \bibinfo {author} {\bibfnamefont {G.~E.}\ \bibnamefont
  {Bauer}},  \emph {et~al.},\ }\href@noop {} {\bibfield  {journal} {\bibinfo
  {journal} {Physical Review X}\ }\textbf {\bibinfo {volume} {12}},\ \bibinfo
  {pages} {011060} (\bibinfo {year} {2022})}\BibitemShut {NoStop}%
\bibitem [{\citenamefont {Saito}\ \emph {et~al.}(2003)\citenamefont {Saito},
  \citenamefont {Matsuda},\ and\ \citenamefont {Wright}}]{saito2003picosecond}%
  \BibitemOpen
  \bibfield  {author} {\bibinfo {author} {\bibfnamefont {T.}~\bibnamefont
  {Saito}}, \bibinfo {author} {\bibfnamefont {O.}~\bibnamefont {Matsuda}}, \
  and\ \bibinfo {author} {\bibfnamefont {O.}~\bibnamefont {Wright}},\
  }\href@noop {} {\bibfield  {journal} {\bibinfo  {journal} {Physical Review
  B}\ }\textbf {\bibinfo {volume} {67}},\ \bibinfo {pages} {205421} (\bibinfo
  {year} {2003})}\BibitemShut {NoStop}%
\bibitem [{\citenamefont {Salikhov}\ \emph {et~al.}(2019)\citenamefont
  {Salikhov}, \citenamefont {Alekhin}, \citenamefont {Parpiiev}, \citenamefont
  {Pezeril}, \citenamefont {Makarov}, \citenamefont {Abrudan}, \citenamefont
  {Meckenstock}, \citenamefont {Radu}, \citenamefont {Farle}, \citenamefont
  {Zabel},\ and\ \citenamefont {Temnov}}]{Salikhov2019}%
  \BibitemOpen
  \bibfield  {author} {\bibinfo {author} {\bibfnamefont {R.}~\bibnamefont
  {Salikhov}}, \bibinfo {author} {\bibfnamefont {A.}~\bibnamefont {Alekhin}},
  \bibinfo {author} {\bibfnamefont {T.}~\bibnamefont {Parpiiev}}, \bibinfo
  {author} {\bibfnamefont {T.}~\bibnamefont {Pezeril}}, \bibinfo {author}
  {\bibfnamefont {D.}~\bibnamefont {Makarov}}, \bibinfo {author} {\bibfnamefont
  {R.}~\bibnamefont {Abrudan}}, \bibinfo {author} {\bibfnamefont
  {R.}~\bibnamefont {Meckenstock}}, \bibinfo {author} {\bibfnamefont
  {F.}~\bibnamefont {Radu}}, \bibinfo {author} {\bibfnamefont {M.}~\bibnamefont
  {Farle}}, \bibinfo {author} {\bibfnamefont {H.}~\bibnamefont {Zabel}}, \ and\
  \bibinfo {author} {\bibfnamefont {V.~V.}\ \bibnamefont {Temnov}},\ }\href
  {\doibase 10.1103/PhysRevB.99.104412} {\bibfield  {journal} {\bibinfo
  {journal} {Phys. Rev. B}\ }\textbf {\bibinfo {volume} {99}},\ \bibinfo
  {pages} {104412} (\bibinfo {year} {2019})}\BibitemShut {NoStop}%
\bibitem [{\citenamefont {Hatanaka}\ \emph {et~al.}(2022)\citenamefont
  {Hatanaka}, \citenamefont {Asano}, \citenamefont {Okamoto}, \citenamefont
  {Kunihashi}, \citenamefont {Sanada},\ and\ \citenamefont
  {Yamaguchi}}]{hatanaka2022chip}%
  \BibitemOpen
  \bibfield  {author} {\bibinfo {author} {\bibfnamefont {D.}~\bibnamefont
  {Hatanaka}}, \bibinfo {author} {\bibfnamefont {M.}~\bibnamefont {Asano}},
  \bibinfo {author} {\bibfnamefont {H.}~\bibnamefont {Okamoto}}, \bibinfo
  {author} {\bibfnamefont {Y.}~\bibnamefont {Kunihashi}}, \bibinfo {author}
  {\bibfnamefont {H.}~\bibnamefont {Sanada}}, \ and\ \bibinfo {author}
  {\bibfnamefont {H.}~\bibnamefont {Yamaguchi}},\ }\href@noop {} {\bibfield
  {journal} {\bibinfo  {journal} {Physical Review Applied}\ }\textbf {\bibinfo
  {volume} {17}},\ \bibinfo {pages} {034024} (\bibinfo {year}
  {2022})}\BibitemShut {NoStop}%
\bibitem [{\citenamefont {Domke}\ \emph {et~al.}(2014)\citenamefont {Domke},
  \citenamefont {Nobile}, \citenamefont {Rapp}, \citenamefont {Eiselen},
  \citenamefont {Sotrop}, \citenamefont {Huber},\ and\ \citenamefont
  {Schmidt}}]{domke2014understanding}%
  \BibitemOpen
  \bibfield  {author} {\bibinfo {author} {\bibfnamefont {M.}~\bibnamefont
  {Domke}}, \bibinfo {author} {\bibfnamefont {L.}~\bibnamefont {Nobile}},
  \bibinfo {author} {\bibfnamefont {S.}~\bibnamefont {Rapp}}, \bibinfo {author}
  {\bibfnamefont {S.}~\bibnamefont {Eiselen}}, \bibinfo {author} {\bibfnamefont
  {J.}~\bibnamefont {Sotrop}}, \bibinfo {author} {\bibfnamefont {H.~P.}\
  \bibnamefont {Huber}}, \ and\ \bibinfo {author} {\bibfnamefont
  {M.}~\bibnamefont {Schmidt}},\ }\href@noop {} {\bibfield  {journal} {\bibinfo
   {journal} {Physics Procedia}\ }\textbf {\bibinfo {volume} {56}},\ \bibinfo
  {pages} {1007} (\bibinfo {year} {2014})}\BibitemShut {NoStop}%
\bibitem [{\citenamefont {Temnov}\ \emph {et~al.}(2020)\citenamefont {Temnov},
  \citenamefont {Alekhin}, \citenamefont {Samokhvalov}, \citenamefont {Ivanov},
  \citenamefont {Lomonosov}, \citenamefont {Vavassori}, \citenamefont {Modin},\
  and\ \citenamefont {Veiko}}]{temnov2020nondestructive}%
  \BibitemOpen
  \bibfield  {author} {\bibinfo {author} {\bibfnamefont {V.~V.}\ \bibnamefont
  {Temnov}}, \bibinfo {author} {\bibfnamefont {A.}~\bibnamefont {Alekhin}},
  \bibinfo {author} {\bibfnamefont {A.}~\bibnamefont {Samokhvalov}}, \bibinfo
  {author} {\bibfnamefont {D.~S.}\ \bibnamefont {Ivanov}}, \bibinfo {author}
  {\bibfnamefont {A.}~\bibnamefont {Lomonosov}}, \bibinfo {author}
  {\bibfnamefont {P.}~\bibnamefont {Vavassori}}, \bibinfo {author}
  {\bibfnamefont {E.}~\bibnamefont {Modin}}, \ and\ \bibinfo {author}
  {\bibfnamefont {V.~P.}\ \bibnamefont {Veiko}},\ }\href@noop {} {\bibfield
  {journal} {\bibinfo  {journal} {Nano Letters}\ }\textbf {\bibinfo {volume}
  {20}},\ \bibinfo {pages} {7912} (\bibinfo {year} {2020})}\BibitemShut
  {NoStop}%
\bibitem [{\citenamefont {Jaris}\ \emph {et~al.}(2020)\citenamefont {Jaris},
  \citenamefont {Yang}, \citenamefont {Berk},\ and\ \citenamefont
  {Schmidt}}]{jaris2020towards}%
  \BibitemOpen
  \bibfield  {author} {\bibinfo {author} {\bibfnamefont {M.}~\bibnamefont
  {Jaris}}, \bibinfo {author} {\bibfnamefont {W.}~\bibnamefont {Yang}},
  \bibinfo {author} {\bibfnamefont {C.}~\bibnamefont {Berk}}, \ and\ \bibinfo
  {author} {\bibfnamefont {H.}~\bibnamefont {Schmidt}},\ }\href@noop {}
  {\bibfield  {journal} {\bibinfo  {journal} {Physical Review B}\ }\textbf
  {\bibinfo {volume} {101}},\ \bibinfo {pages} {214421} (\bibinfo {year}
  {2020})}\BibitemShut {NoStop}%
\bibitem [{\citenamefont {Yang}\ and\ \citenamefont
  {Schmidt}(2021)}]{yang2021acoustic}%
  \BibitemOpen
  \bibfield  {author} {\bibinfo {author} {\bibfnamefont {W.-G.}\ \bibnamefont
  {Yang}}\ and\ \bibinfo {author} {\bibfnamefont {H.}~\bibnamefont {Schmidt}},\
  }\href@noop {} {\bibfield  {journal} {\bibinfo  {journal} {Applied Physics
  Reviews}\ }\textbf {\bibinfo {volume} {8}},\ \bibinfo {pages} {021304}
  (\bibinfo {year} {2021})}\BibitemShut {NoStop}%
\bibitem [{\citenamefont {Berger}\ \emph {et~al.}(2015)\citenamefont {Berger},
  \citenamefont {de~la Osa}, \citenamefont {Suszka}, \citenamefont {Pancaldi},
  \citenamefont {S{\'a}iz}, \citenamefont {Moreno}, \citenamefont {Oepen},\
  and\ \citenamefont {Vavassori}}]{berger2015enhanced}%
  \BibitemOpen
  \bibfield  {author} {\bibinfo {author} {\bibfnamefont {A.}~\bibnamefont
  {Berger}}, \bibinfo {author} {\bibfnamefont {R.~A.}\ \bibnamefont {de~la
  Osa}}, \bibinfo {author} {\bibfnamefont {A.~K.}\ \bibnamefont {Suszka}},
  \bibinfo {author} {\bibfnamefont {M.}~\bibnamefont {Pancaldi}}, \bibinfo
  {author} {\bibfnamefont {J.~M.}\ \bibnamefont {S{\'a}iz}}, \bibinfo {author}
  {\bibfnamefont {F.}~\bibnamefont {Moreno}}, \bibinfo {author} {\bibfnamefont
  {H.~P.}\ \bibnamefont {Oepen}}, \ and\ \bibinfo {author} {\bibfnamefont
  {P.}~\bibnamefont {Vavassori}},\ }\href@noop {} {\bibfield  {journal}
  {\bibinfo  {journal} {Physical review letters}\ }\textbf {\bibinfo {volume}
  {115}},\ \bibinfo {pages} {187403} (\bibinfo {year} {2015})}\BibitemShut
  {NoStop}%
\end{thebibliography}

%

\end{document}